\documentclass[journal,twoside,web]{ieeecolor}
\usepackage{tmi}
\usepackage{cite}
\usepackage{amsmath,amssymb,amsfonts}
\usepackage{mathtools}
\usepackage{algorithmic}
\usepackage{graphicx}
\usepackage{textcomp}
\usepackage{epsfig}
\usepackage{booktabs}
\usepackage{multirow}
\usepackage{hyperref}
\usepackage{url}
\usepackage{tabularx}
\usepackage{subfigure}
\usepackage{xcolor}
\hypersetup{
    colorlinks=true,
    linkcolor=blue,
    filecolor=magenta,
    urlcolor=cyan,
}
\usepackage{color}

\def\BibTeX{{\rm B\kern-.05em{\sc i\kern-.025em b}\kern-.08em
    T\kern-.1667em\lower.7ex\hbox{E}\kern-.125emX}}
\begin{document}
\title{Contralaterally Enhanced Networks for \\ Thoracic Disease Detection}
\author{Gangming Zhao,
        Chaowei Fang,
        Guanbin Li, \IEEEmembership{Member, IEEE}, \\
        Licheng Jiao, \IEEEmembership{Fellow, IEEE},
        and Yizhou Yu, \IEEEmembership{Fellow, IEEE},
        \thanks{\textit{Gangming Zhao and Chaowei Fang contributed equally to this work.}}
\thanks{G. Zhao is with the Department of Computer Science, The University of Hong Kong, Hong Kong.}
\thanks{C. Fang and L. Jiao are with the School of Artificial Intelligence, Xidian University, No. 2 South Taibai Road, Xi'an, Shaanxi, 710071, China (e-mail: cwfang@xidian.edu.cn, lchjiao@mail.xidian.edu.cn).}
\thanks{G. Li is with the School of Data and Computer Science, Sun Yat-Sen University, 510006, China (e-mail: liguanbin@mail.sysu.edu.cn).}
\thanks{Y. Yu is with DeepWise AI Lab, Beijing, 100080, China (e-mail: yizhouy@acm.org).}
}

\maketitle

\begin{abstract}
Identifying and locating diseases in chest X-rays are very challenging, due to the low visual contrast between normal and abnormal regions, and distortions caused by other overlapping tissues.
An interesting phenomenon is that there exist many similar structures in the left and right parts of the chest, such as ribs, lung fields and bronchial tubes.  This kind of similarities can be used to identify diseases in chest X-rays, according to the experience of broad-certificated radiologists.
Aimed at improving the performance of existing detection methods, we propose a deep end-to-end module to exploit the contralateral context information for enhancing feature representations of disease proposals.
First of all, under the guidance of the spine line, the spatial transformer network is employed to extract local contralateral patches, which can provide valuable context information for disease proposals.
Then, we build up a specific module, based on both additive and subtractive operations, to fuse the features of the disease proposal and the contralateral patch.
Our method can be integrated into both fully and weakly supervised disease detection frameworks.
It achieves 33.17 AP50 on a carefully annotated private chest X-ray dataset which contains 31,000 images.
Experiments on the NIH chest X-ray dataset indicate that our method achieves state-of-the-art performance in weakly-supervised disease localization.

\end{abstract}


\section{Introduction}
\IEEEPARstart{C}{hest} X-ray (CXR) is one of the most widely-used examination tools for the diagnosis of thoracic diseases such as lung nodules and pneumonia. Thanks to the development of deep learning technologies, stupendous progress has been achieved in automatic disease classification~\cite{wang2017chestx,aviles2019graphx} and localization~\cite{cai2018iterative,li2018thoracic,liu2019align} for chest X-rays. Considering there exist similar structures in the left and right parts of the chest, we focus on exploring the contralateral context information for both fully and weakly supervised disease detection.

The main challenges of disease identification in chest X-ray images include low visual contrast between lesion regions and other components, and distortions induced by other overlapping tissues. Sometimes it is difficult for medical specialists to recognize obscure diseases~\cite{rajpurkar2017chexnet,Qin2018}. Designing automatic artificial intelligence systems is beneficial for guaranteeing the diagnosis efficiency and accuracy. Previous deep learning methods for chest X-ray diagnosis mainly concentrated on disease classification~\cite{Qin2018,brestel2018radbot,irvin2019chexpert,aviles2019graphx}. Recently, several literatures researched on detecting disease regions under weak/limited supervision. They can be grouped into two main categories: the first category of methods~\cite{wang2017chestx,cai2018iterative} resort to convolutional neural networks (CNN) trained on the classification task and output disease localization results through calculating category activation maps~\cite{zhou2016learning}; the second kind of methods~\cite{li2018thoracic,liu2019align} use the multiple instance learning to directly yield categoric probability maps which can be easily transformed into lesion detections. However, the performance of these methods is still far from practical clinical usage.

\begin{figure}[t]
\begin{center}
\includegraphics[width=1\linewidth]{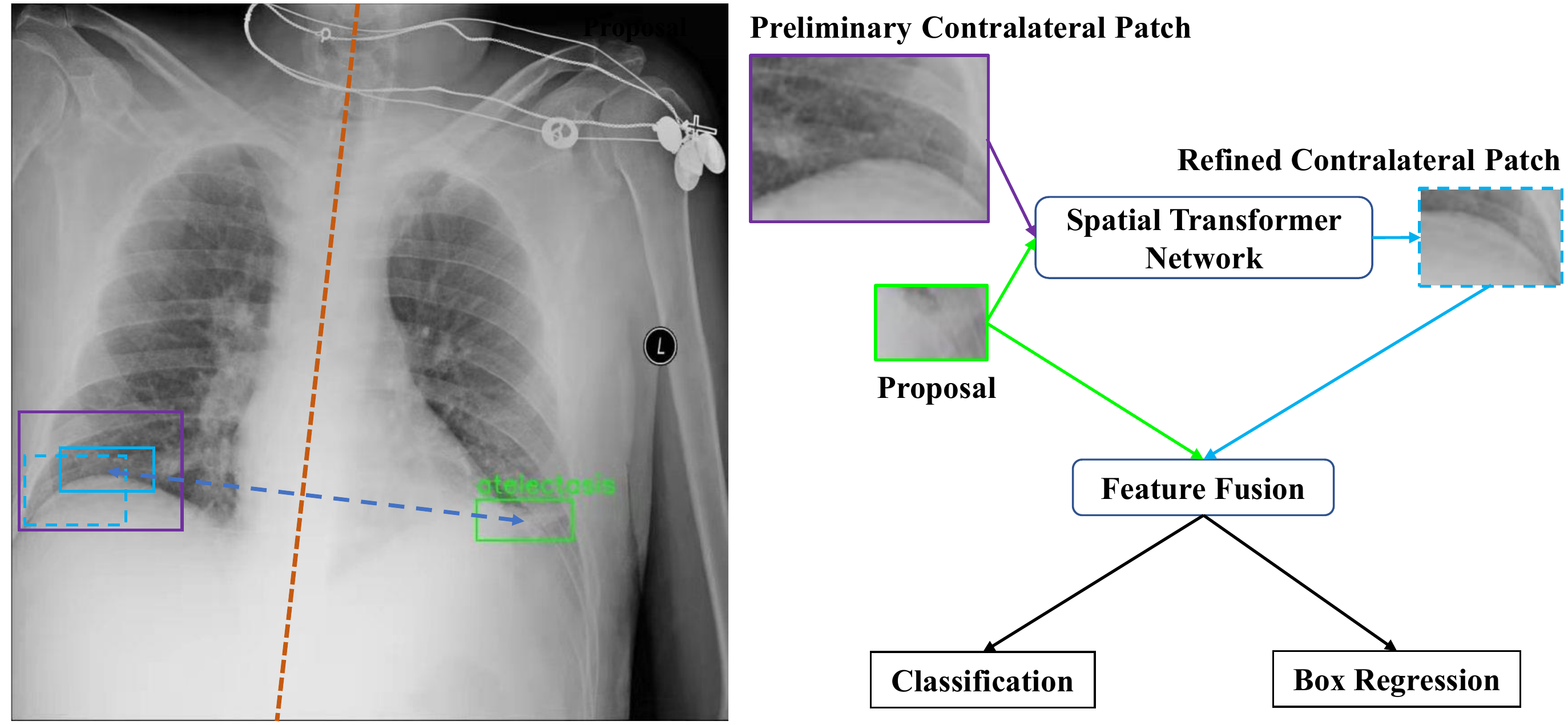}
\end{center}
\caption{For each disease proposal, we seek its initial contralateral patch under the guidance of the spine line. Then its pose is automatically adjusted to acquire a more appropriate patch to facilitate the final classification and localization of the disease proposal.}\label{fig:intro}
\end{figure}

A mainstream pipeline of object detection is to screen out potential proposals followed by identifying the class of proposals~\cite{ren2015faster,yan2018deeplesion}. Through stacking piles of convolutional layers, CNN models are very advantageous at extracting surrounding context information. However, distant relationships are still hard to be exploited with convolutions which usually have small kernels. Particularly for chest X-rays, the left and right parts of the chest share lots of similar structures, such as ribs, lung fields and bronchial tubes. Although the two halves of the chest are not symmetrical (e.g., the left and right lung is composed of 2 and 3 lobes respectively, and the heart resides only in the left side of the chest), we are wondering whether the similarity information can exert a positive influence on disease detection. As shown in Fig. \ref{fig:intro}, we devise a novel CNN-based module, named \textit{Contralaterally Enhanced Networks}, to take advantage of such similarity information.  First of all, we employ existing methods to acquire a number of disease proposals from the input image. For the sake of extracting the similarity information from the contralateral part of the chest, the spine line is used as the symmetrical axis to obtain a reference patch for every disease proposal. Then, with the help of the spatial transformer network~\cite{jaderberg2015spatial}, we sample an appropriate patch from a relatively large region enclosing the contralateral reference patch, to enhance the feature representation of the disease proposal. Finally, a fusion module is devised to aggregate the features of the disease proposal and its contralateral reference patch. Our proposed method can be plugged into both fully and weakly supervised disease detection frameworks. The main contributions of this paper can be summarized as follows.
\begin{itemize}
    \item We build up a novel deep module, \textit{Contralaterally Enhanced Networks}, for facilitating the disease detection in chest X-rays. We are the first to explicitly exploit the contralateral context information between the left and right parts of chest, to enhance the feature representations of disease proposals.
    \item An effective method is proposed to seek contralateral reference patches for disease proposals. A reference patch is extracted for each disease proposal under the guidance of the spine line and is further refined with the spatial transformer network. And A novel feature fusion module is devised to enhance the feature representation of a disease proposal with its contralateral reference patch.
    \item Our proposed method can improve existing object detection baselines~\cite{ren2015faster,lin2017feature,lin2017focal,duan2019centernet} with large margins on a chest X-ray dataset for fully supervised disease detection. It also achieves state-of-the-art performance on the NIH chest X-ray dataset~\cite{wang2017chestx} under the weakly supervised setting.
\end{itemize}

\section{Related Work}
\subsection{Object Detection}
Object detection is a widely-studied topic in both natural and medical images. It aims at localizing object instances of interest such as faces, pedestrians and disease lesions. The most famous kind of deep learning approaches for object detection is the R-CNN~\cite{girshick2014rich} family. The primitive R-CNN extracts proposals through selective search~\cite{uijlings2013selective}, and then predicts object bounding boxes and  categories from convolution features of these proposals. Fast R-CNN~\cite{girshick2015fast} adopts a shared backbone network to extract proposal features via RoI pooling. Faster R-CNN~\cite{ren2015faster} automatically produces object proposals from top-level features with the help of pre-defined anchors. The above methods accomplish the detection procedure through two stages, including object proposal extraction, object recognition and localization. In~\cite{lin2017feature}, the feature pyramid network is exploited to further improve the detection performance of Fast R-CNN and Faster R-CNN with the help of multi-scale feature maps.
The other pipeline for object detection implements object localization and identification in single stage through simultaneous bounding box regression and object classification, such as YOLO~\cite{redmon2016you} and SSD~\cite{liu2016ssd}. The RetinaNet~\cite{lin2017focal} is also built upon the feature pyramid network, and uses dense box predictions during the training stage. The focal loss is proposed to cope with the class imbalance problem. Reference~\cite{law2018cornernet} presents an anchor-free pipeline through detecting corners of bounding boxes. Despite of corners, the center point is also explored to guarantee the correctness of the obtained object boxes in CenterNet~\cite{duan2019centernet}.
The detection task has also attracted a large amount of research interest in medical images, such as lesion detection in CT scans~\cite{yan2018deeplesion} and cell detection in malaria images~\cite{hung2017applying}. This paper targets at detecting diseases in chest X-ray images. Practically, we propose a \textit{Contralaterally Enhanced Networks} to exploit contralateral context information to enhance feature representations of disease proposals.

\subsection{Disease Detection in Chest X-ray Images}
Accurately recognizing and localizing diseases in chest X-Ray images is very challenging because of low textural contrast, large anatomic variation across patients, and organ overlapping. Previous works in this field mainly focus on disease classification~\cite{noor2014texture,cicero2017training,rajpurkar2017chexnet,wang2017chestx,aviles2019graphx}. Recently, the authors in~\cite{chouhan2020novel} propose to transfer deep models pretrained on the ImageNet dataset~\cite{deng2009imagenet} for recognizing pneumonia in chest X-ray images. In~\cite{sahlol2020novel}, the artificial ecosystem-based optimization algorithm is used to select the most relevant features for tuberculosis recognition.
Based on the category activation map~\cite{zhou2016learning} which can be estimated with a disease recognition network, researchers attempt to localize disease in a weakly supervised manner~\cite{wang2017chestx,cai2018iterative}.
In~\cite{zhang2020thoracic}, the triplet loss is used to facilitate the training of the disease classification model, and better performance is observed in class activation maps (CAM) estimated by the trained model.
In~\cite{li2018thoracic,liu2019align}, multiple instance learning is employed to solve the disease localization problem.
In~\cite{zhou2018weakly}, a novel weakly supervised disease detection model is devised on the basis of the DenseNet~\cite{huang2017densely}. Two pooling layers including a class-wise pooling layer and a spatial pooling layer are used to transform 2-dimensional class attention maps into the final prediction scores. The performance of these methods is still far from practical usage in automatic diagnosis systems.
In~\cite{ke2019neuro}, a novel pipeline is proposed to identify and search potential lung diseases with the help of heuristic algorithms, such as Moth-Flame and Ant Lion.

An interesting phenomenon is that there exist similar structures between the left and right halves of the chest, such as the left and right parts of lungs. Such contralateral context information can benefit the recognition of thoracic diseases according to the experience of broad-certificated radiologists. Reference~\cite{santosh2017automated} attempts to take advantage of the lung region symmetry when constructing hand-crafted features for thoracic disease or abnormality identification.
In this paper, we devise a specific module to extract context information from the contralateral structures for strengthening the feature representations of disease proposals. The contralateral context information can effectively improve the performance of existing fully and weakly supervised disease detection methods.

\begin{figure*}[t]
\begin{center}
\includegraphics[width=1\linewidth]{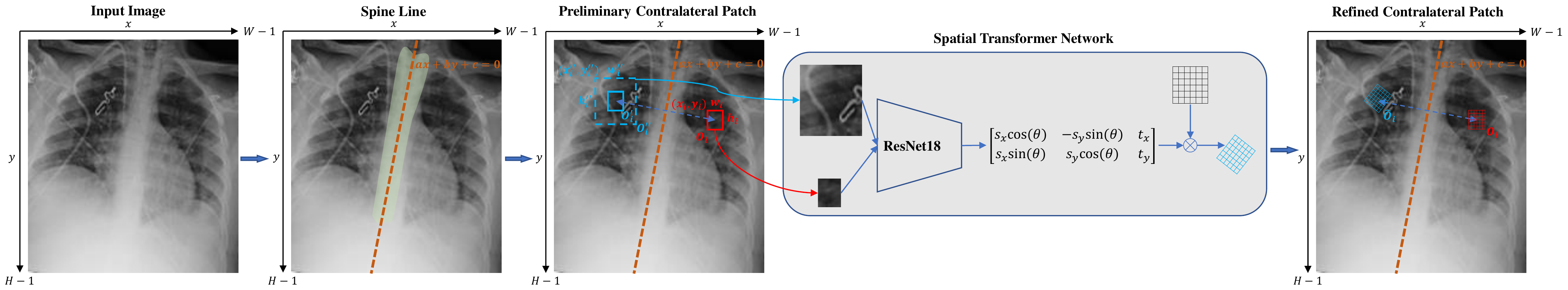}
\end{center}
\caption{Procedure of extracting the contralateral patch for each disease proposal. First the spine line is derived from the spine mask. For each disease proposal $O_i$, a preliminary contralateral patch $O_i^\prime$ is retrieved under the guidance of the spine line. Then a spatial transformer network is devised to acquire the final contralateral patch $\hat O_i$. } \label{fig:symmpatch}
\end{figure*}

\begin{figure}[t]
\begin{center}
\includegraphics[width=1\linewidth]{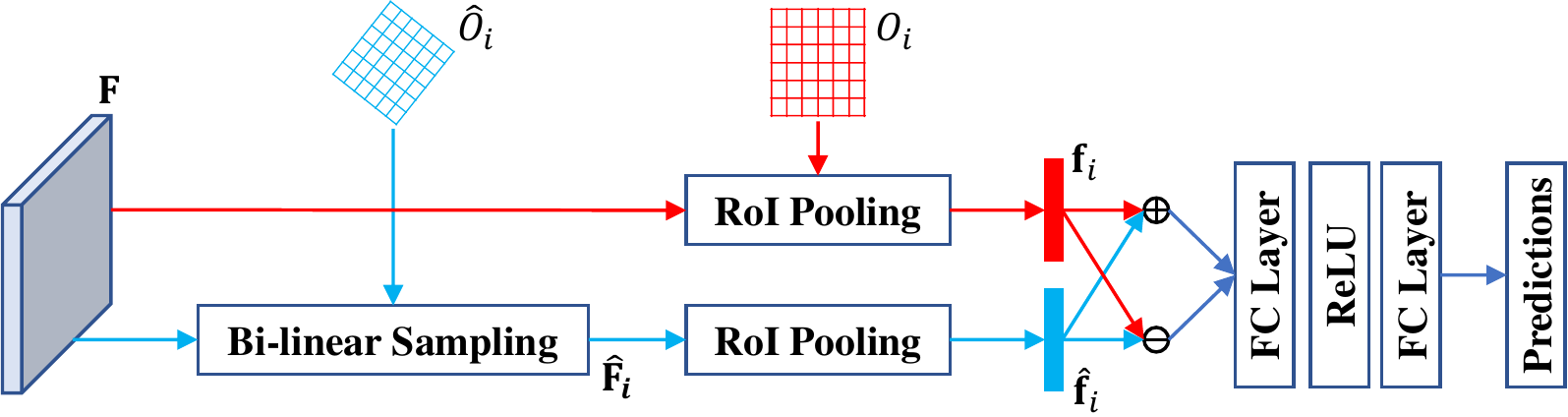}
\end{center}
\caption{Feature fusion module. For the disease proposal $O_i$, ROI pooling is directly used to extract the feature representation from the input feature map $\mathbf F$. For the contralateral patch $\hat O_i$, a specific feature map is generated via bi-linear sampling, which is then used to produce a feature representation using ROI pooling as well. The feature representations of $O_i$ and $\hat O_i$ are fused, and then fed into two fully connected layers to produce the final predictions. } \label{fig:fusion}
\end{figure}

\section{Contralaterally Enhanced Networks} \label{sec:cen}
\iffalse
\begin{figure}[t]
\begin{center}
\includegraphics[width=1\columnwidth]{figures/splinesegnet.pdf}
\end{center}
\caption{The network architecture for extracting spline line. We use PSPNet as the model for segmenting the spline mask at first. The spline line is then located as the central line of the mask.}
\end{figure}
\fi

The target of this paper is to automatically locate various thoracic diseases in chest X-rays. There exists a certain degree of similarity between the left and right parts of the chest, from high-level structures of organs, such as lungs, bones and vessels, to low-level tissues. Based on this observation, we propose a feature enhancement module to exploit the contralateral context information for enhancing the feature representations of disease proposals.
For every disease proposal extracted with an existing disease detection method, a reference patch from the contralateral location of the chest is acquired under the guidance of the spine line at first. Then a transformer network is devised to refine the pose of the reference patch, which is used to complement the representation of the disease proposal via an additive and subtractive feature fusion module.  Our proposed module can be easily integrated into both fully and weakly supervised disease detection models. The technical details are illustrated as below.

\subsection{Contralateral Patch Extraction}
Given a CXR image $I$ with size of $W\times H$, we can screen out $n$ potential disease proposals $\{O_i | i=1\cdots n\}$ with the help of an existing disease/object detection method, such as the fully-supervised method \cite{ren2015faster} or the weakly-supervised method \cite{liu2019align}. $O_i$ is represented by a quad $(x_i, y_i, w_i, h_i)$, indicating the horizontal and vertical coordinate of the top-left corner, width, and height, respectively. We denote the disease category of $O_i$ as $l_i$. Suppose the number of target disease categories is $m$. Hence, $l_i\in\{1,\cdots,m\}$. The pipeline of extracting the contralateral patch for each disease proposal is illustrated in Fig. \ref{fig:symmpatch}.

\subsubsection{Preliminary Contralateral Patch}
Considering the spine is located at a relatively middle position of the chest, we exploit the spine line to fetch the preliminary contralateral patch for each disease proposal. The minimum circumscribed quadrilateral enclosing the spine mask can be obtained as in~\cite{eppstein1992finding}. We regard the spine line bridged by the centers of two short edges as the symmetric axis, which can be expressed as $ax+by+c=0$ ($a$, $b$ and $c$ are coefficients). For a disease proposal $O_i$, its preliminary contralateral patch $O_i^\prime$($=(x_i^\prime,y_i^\prime,w_i,h_i)$) is located through solving the following linear system,
\begin{equation}
\begin{cases}
a\frac{x_i+x_i^\prime+w_i-1}{2}+ b\frac{y_i+y_i^\prime+h_i-1}{2}+c=0,\\
-b(x_i-x_i^\prime)+a(y_i-y_i^\prime)=0.
\end{cases}
\end{equation}

The PSPNet proposed in~\cite{zhao2017pyramid} is chosen as our spine segmentation model. We use ResNet50~\cite{he2016deep} as the backbone of PSPNet and modify the dimension of the final output into 1. The same settings as in~\cite{zhao2017pyramid} are adopted to optimize the network parameters. Quantitative and qualitive experimental results are reported in Section \ref{sec:exper-full}.

\begin{figure*}[t]
\center
\includegraphics[width=0.9\linewidth]{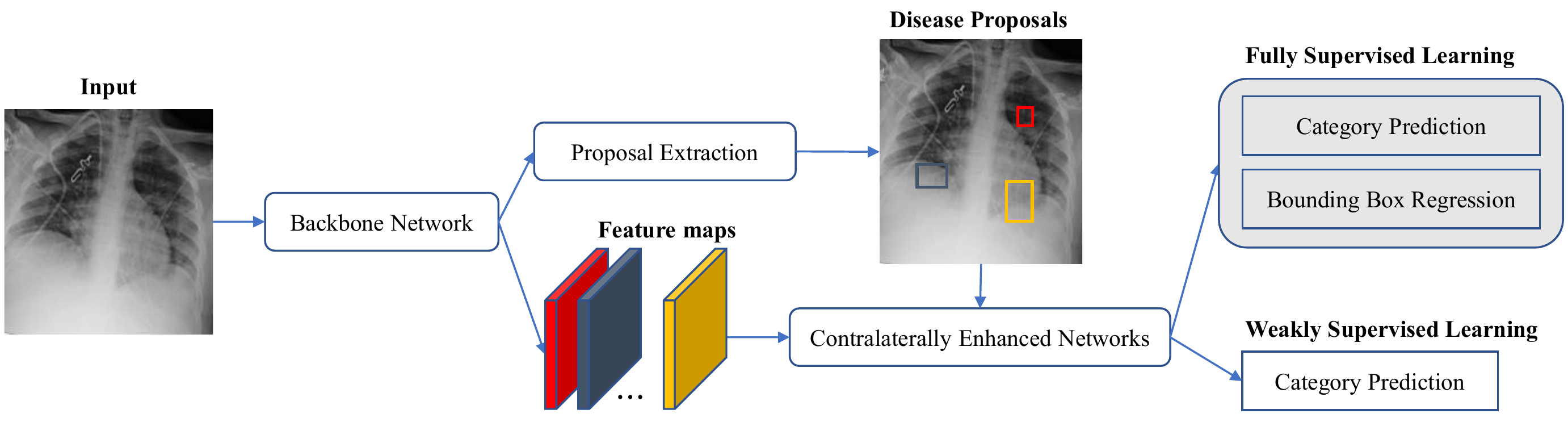}
\caption{Disease detection frameworks. The contralaterally enhanced networks can be integrated into both fully and weakly supervised methods.} \label{fig:frame}
\end{figure*}
\subsubsection{Refined Contralateral Patch}
To acquire a patch which is more suitable for enhancing the feature representation of the disease proposal, we further devise a spatial transformer network (abbr. STN) to refine the pose of $O_i^\prime$. In details, we set up STN on the basis of ResNet18~\cite{he2016deep}, through modifying the output dimension of the ultimate fully connected layer into 6. %\new{Note that, the ResNet18 model is pretrained on ImageNet dataset[].}
 We extend the borders of $O_i^\prime$ by $\Delta_x$ and $\Delta_y$ pixels along the horizontal and vertical axes respectively, resulting to a new patch $O_i^{\prime\prime}=(x_i^{\prime\prime}, y_i^{\prime\prime}, w_i^{\prime\prime}, h_i^{\prime\prime})$. $x_i^{\prime\prime}=x_i^\prime-\Delta_x$, $y_i^{\prime\prime}=y_i^\prime-\Delta_y$, $w_i^{\prime\prime}=w_i+2\Delta_x$, and $h_i^{\prime\prime}=h_i+2\Delta_y$. Empirically, we set $\Delta_x=0.25w_i$ and $\Delta_y=0.25h_i$.  Afterwards the original disease proposal $O_i$ is padded to have the same size as $O_i^{\prime\prime}$. After being resized to a fixed size of $w^0\times h^0$, the padded disease proposal and $O_i^{\prime\prime}$ are concatenated and fed into the STN, which gives rise to a tensor of 6 elements including two rescaling parameters ($s_x$ and $s_y$), two transition parameters ($t_x$ and $t_y$) and one rotation parameter ($\theta$). These parameters can be used to locate the refined symmetrical patch $\hat O_i$. Here, $w^0$ and $h^0$ are both set as 64.  For a point $(x^\textrm{dst}, y^\textrm{dst})$ in $\hat O_i$, we can obtain its corresponding point $(x^\textrm{src}, y^\textrm{src})$ in the original input image with a linear transformation operation, $[x^\textrm{dst}, y^\textrm{dst}]^\textrm{T}=\mathbf T_i [x^\textrm{src}, y^\textrm{src}, 1]^\textrm{T}$. $\mathbf T_i$ can be calculated as the following formulation,
\begin{equation} \label{eq:trans}
\mathbf T_i=
\begin{bmatrix}
  \frac{w^{\prime\prime}_i-1}{w_0-1} & 0  & x_i^{\prime\prime} \\
  0 & \frac{h^{\prime\prime}_i-1}{h_0-1}  & y_i^{\prime\prime} \\
\end{bmatrix}
\begin{bmatrix}
  s_x\cos\theta & -s_y\sin\theta & t_x \\
  s_x\sin\theta & s_y\cos\theta  & t_y \\
  0 & 0 & 1 \\
\end{bmatrix}.
\end{equation}

\subsection{Feature Fusion Module}
%($\lfloor \frac{W_f}{W}w_i \rfloor \times \lfloor \frac{H_f}{H}h_i \rfloor$)
The architecture of the module for fusing features of a disease proposal and its contralateral patch is shown in Fig. \ref{fig:fusion}. Given a feature map $\mathbf F$ extracted with a backbone network, the RoI pooling operation~\cite{girshick2015fast} is adopted to extract feature representation $\mathbf f_i$ for the disease proposal $O_i$. %Suppose the width and height of $\mathbf F$ be $W_f$ and $H_f$, respectively.
A feature map $\hat{\mathbf F}_i$ ($w\times h$) is extracted for the contralateral patch $\hat O_i$. For every point in $\hat O_i$, its coordinates are obtained according to transformation matrix in (\ref{eq:trans}), and its feature vector is sampled from from $\mathbf F$ via bi-linear interpolation.
%through bi-linear interpolation.  %\new{The bi-linear interpolation is also adopted in [] and [] et al. Since our STN transforms the regular grids to be irregular, the operation is usually used to acquire the value of the irregular points.}
%Every point in $\hat{\mathbf F}_i$ can be easily located in $\mathbf F$ with the help of the transformation operation (\ref{eq:trans}).
Again RoI pooling is used to aggregate $\hat{\mathbf F}_i$ into the feature representation $\mathbf{\hat f}_i$ for $\hat O_i$. The spatial sizes of $\mathbf f_i$ and $\mathbf{\hat f}_i$ are both set as $7\times 7$.  We directly flatten $\mathbf f_i$ and $\mathbf{\hat f}_i$ into 1-dimensional feature vectors. Every element in $\mathbf f_i$ or $\mathbf{\hat f}_i$ can be regarded as certain attribute factor of $O_i$ or $\hat O_i$. The addition between $\mathbf f_i$ and $\mathbf{\hat f}_i$ can help highlight attributes which have large responses in both $O_i$ and $\hat O_i$. This is beneficial to the identification of diseases striding across the left and right parts of the chest. On the other hand, the subtraction between $\mathbf f_i$ and $\mathbf{\hat f}_i$ can provide contrast information and suppress the responses of attributes which are irrelevant to disease recognition and localization. Considering the above issues, both addition and subtraction operations are used to merge $\mathbf f_i$ and $\mathbf{\hat f}_i$. Two fully connected layers are employed to transform the merged features into the final prediction. The output dimension of the first fully connected layer is 512, and that of the second fully connected layer depends on the length of the final prediction.

For fully supervised disease detection, the output is an $m$-dimensional category probability vector $\mathbf p_i$ and 4 $m$-dimensional bounding box offsets including $\mathbf d_i^x$, $\mathbf d_i^y$, $\mathbf d_i^w$ and $\mathbf d_i^h$. Following the parameterizations in~\cite{girshick2014rich}, the updated bounding box for $O_i$ is as bellow,
\begin{eqnarray}
x'_i=w_i*d_i^x(j^\ast)+x_i,& y'_i=h_i*d_i^y(j^\ast)+y_i, \\
w'_i=w_i*e^{d_i^w(j^\ast)},& h'_i=h_i*e^{d_i^h(j^\ast)},
\end{eqnarray}
where $j^\ast=\arg\max_j p_i(j)$. With the help of the above parameterizations, our method can resize the bounding box and translate the top-left corner of the disease proposal.

For weakly supervised disease detection, our method only rectifies the category prediction, and the final output is an $m$-dimensional category probability vector $\mathbf p_i$.
%\new{Specifically, the output dimension of the last fully connected layer is depended on its tasks including classification and localization. If the FC-layer aims for classification objects, its output dimension is 31, otherwise the output dimension is 31 $\times$ 4, where 4 means the top-left point (x, y) of boxes, and the width and height of boxes.}
%The calculation procedure of our core modules including the spatial transformer network and the feature fusion module are differentiable. Thus, no additional supervision information is required for obtaining the contralateral patch.
\\

\begin{table*}[t]
\centering
\caption{Chest X-ray dataset for fully supervised disease detection.}
\label{tab:dataset}

\setlength\tabcolsep{4pt}
\renewcommand{\arraystretch}{1.2}
\begin{tabular}{c|cccccc}\specialrule{.1em}{0em}{0em}
Disease       & aorta widen  & aorta calcification & papilla                & multiple nodules & foreign matter & diffusive nodule        \\
Images/Boxes  & 1,800/6,635  & 1,123/5,144         & 1,210/5,431            & 1,450/4,990      & 1,397/5,127    & 1,240/4,812              \\ \hline
Disease       & cardiomegaly & pneumothorax & hydropneumothorax   & subcutaneous emphysema & cavity           & fibrosis         \\
Images/Boxes  & 1,130/3,125  & 1,120/6,147  & 1,370/6,000         & 1,700/7,101            & 1,100/6,113      & 1,700/8,100      \\ \hline
Disease       & widened mediastinal   & nodule     & rib abnormity& rib absence      & shoulder abnormal      & atelectasis         \\
Images/Boxes  & 1,640/6,135  & 1,200/4,700  & 1,500/5,000  & 1,400/6,100         & 1,727/6,237            & 1,124/4,716          \\ \hline
Disease       & consolidation  & emphysema    & pulmonary tuberculosis & hilum increase & mass & effusion  \\
Images/Boxes  & 1,175/5,764    & 1,034/4,100  & 1,204/6,257            & 1,027/6,131    & 810/3,102 & 910/4,105  \\ \hline
Disease       & pleural thickening  & scoliosis    & subphrenic air    & diaphragm abnormity & calcification & rib fracture  \\
Images/Boxes  & 2,130/8,170         & 814/3,204    & 805/4,017         & 827/3,015    & 723/2,113 & 1,240/7,154  \\ \specialrule{.1em}{0em}{0em}

\iffalse
Disease       & aorta widen & aorta calcification & papilla     & multiple nodules & foreign matter & diffusive nodule & cardiomegaly & pneumothorax  \\
Images/Boxes  & 1,800/6,635 & 1,123/5,144         & 1,210/5,431 & 1,450/4,990      & 1,397/5,127    & 1,240/4,812      & 1,130/3,125  & 1,120/6,147   \\ \hline
Disease       & aorta widen & aorta calcification & papilla     & multiple nodules & foreign matter & diffusive nodule & cardiomegaly & pneumothorax  \\
Images/Boxes  & 1,800/6,635 & 1,123/5,144         & 1,210/5,431 & 1,450/4,990      & 1,397/5,127    & 1,240/4,812      & 1,130/3,125  & 1,120/6,147   \\ \hline
Disease       & aorta widen & aorta calcification & papilla     & multiple nodules & foreign matter & diffusive nodule & cardiomegaly & pneumothorax  \\
Images/Boxes  & 1,800/6,635 & 1,123/5,144         & 1,210/5,431 & 1,450/4,990      & 1,397/5,127    & 1,240/4,812      & 1,130/3,125  & 1,120/6,147   \\ \hline
\fi
\end{tabular}
\end{table*}

\section{Disease Detection Framework}
The overall pipeline for fully and weakly supervised disease detection frameworks is summarized in Fig.~\ref{fig:frame}. Without specification, ResNet50 is used as the backbone of disease detection networks.
\begin{figure*}[t]
%\begin{left}
\centering
\includegraphics[width=0.49\linewidth]{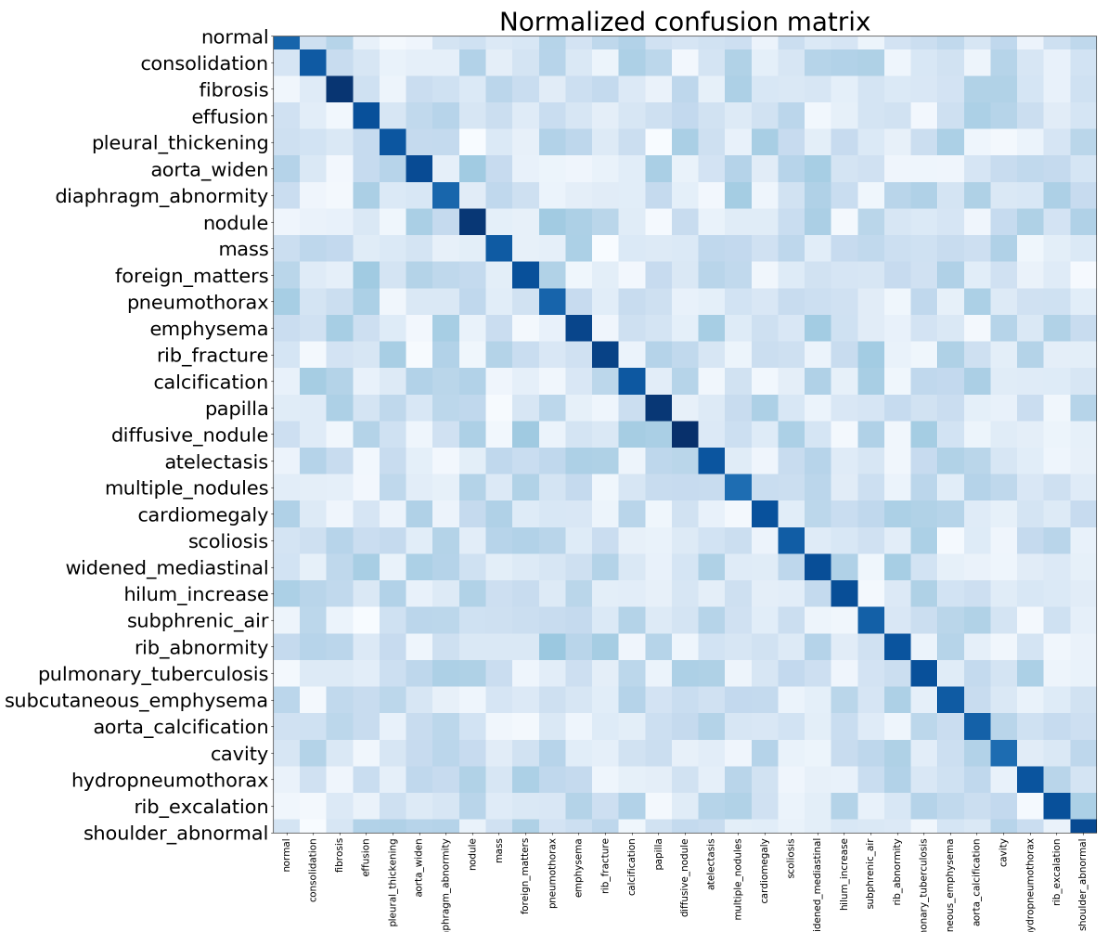}
\includegraphics[width=0.49\linewidth]{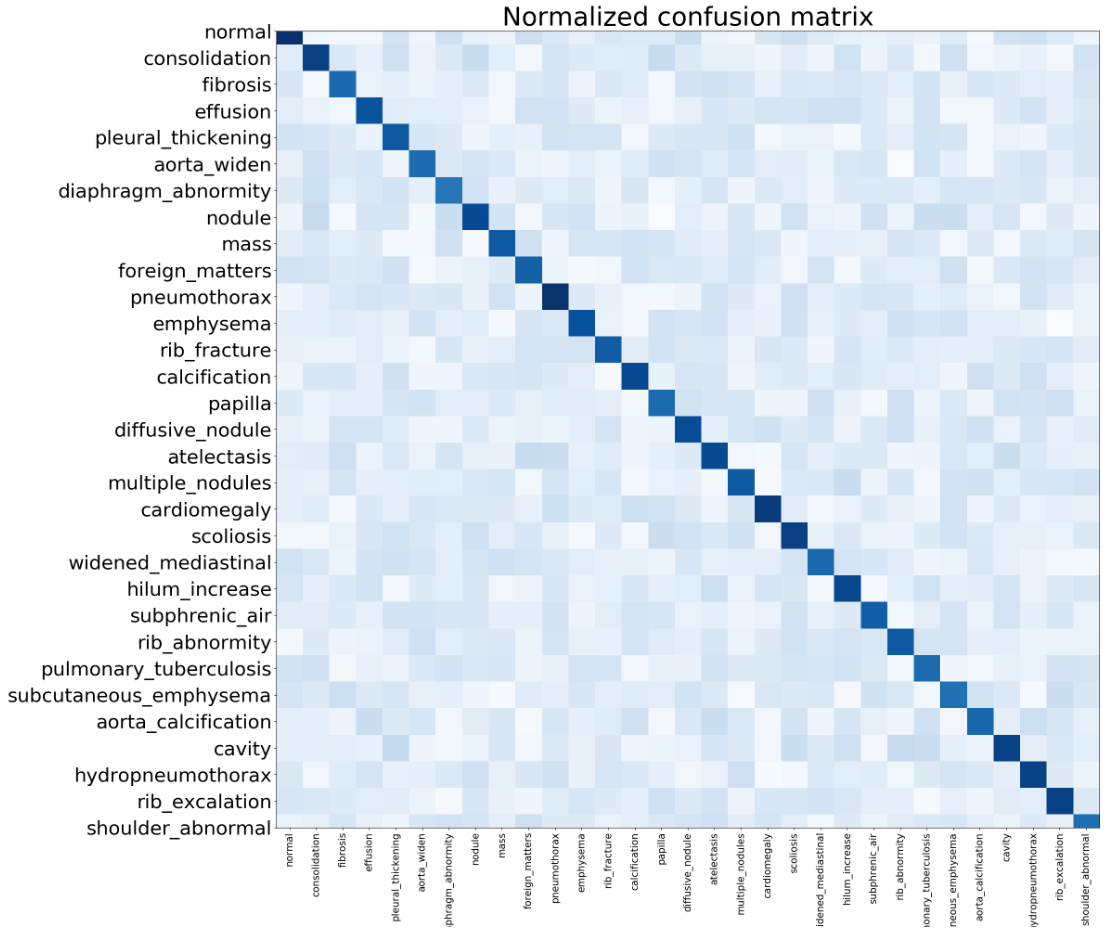}
\begin{tabular}{p{0.55\linewidth}<{\centering}p{0.45\linewidth}<{\centering}}
(a) Vanilla & (b) Ours
\end{tabular}
%\end{left}
\caption{Confusion matrices of disease detection on our private chest X-ray dataset: (a) vanilla Faster RCNN FPN; (b) the variant of Faster RCNN FPN improved by our method. A prediction successfully hits a ground-truth annotation if their classes are consistent and the IoU is larger than 0.5. It is clearly observed that our method significantly reduces false detections. }\label{fig:confusion}
\end{figure*}

%\begin{figure*}[t]
%\begin{left}

%\end{left}
%\caption{\new{The Confusion Matrix of Our Improved Faster R-CNN FPN}}\label{fig:ourcm}
%\end{figure}
\subsection{Fully Supervised Disease Detection} \label{sec:fsd}
Our method can be incorporated into existing fully supervised disease detection frameworks, including both two-stage methods such as Faster R-CNN~\cite{ren2015faster} and Faster R-CNN FPN~\cite{lin2017feature}, and one-stage methods such as RetinaNet~\cite{lin2017focal} and CenterNet~\cite{duan2019centernet}. For two-stage methods, we directly replace the head for predicting the category and bounding box with our contralaterally enhanced networks. It is implemented through feeding the feature map produced by the backbone and class-agnostic disease proposals produced by the region proposal network into our proposed module. When incorporated with one-stage methods, we use the original detection models to produce disease proposals, ignoring the categorical information. Then, our module is adopted to predict the final disease class and rectify the bounding box for every disease proposal. The corresponding feature map which directly induces to the disease proposal is chosen to compute the feature representation for it and its contralateral patch. For the primitive version of Faster R-CNN, the output of the 4-th convolution block (abbr. C4) is adopted as the input feature map of the contralaterally enhanced networks; for models using backbones with pyramid architectures (including Faster R-CNN FPN and RetinaNet), the exact feature map which gives rise to the disease proposal is chosen; for CenterNet, the output of the penultimate convolution layer is used as the input feature map.

We follow the loss function and the training algorithm in \cite{ren2015faster} to optimize our improved disease detection networks.
The loss functions for training the object detection networks
During the testing stage, 100 boxes with the highest confidences are selected as disease proposals, and non-maximum suppression with an IoU threshold of 0.7 is used to filter out severely overlapped boxes. The final predictions are post-processed with the non-maximum suppression again. The IoU threshold is set as 0.5. Boxes with confidences larger than 0.05 are considered as positive detections and the maximum number of boxes is set to 20.

\subsection{Weakly Supervised Disease Detection}
We can also integrate the contralaterally enhanced networks into the weakly supervised disease framework~\cite{liu2019align} which is trained with multiple instance learning (abbr. MIL).  A probability map $\mathbf P$ with size of $\frac{H}{32}\times\frac{W}{32}$ is produced. The vector at position $(x,y)$ in $\mathbf P$ indicate the probabilities of the patch $(32x, 32y, 32, 32)$ with respect to disease categories. Hence, we select top 10 patches as disease proposals and use the output of the penultimate convolutional layer as the feature map which is a tensor having spatial size of $\frac{H}{32}\times\frac{W}{32}$. They are fed into the contralaterally enhanced networks, producing new category probabilities. During the training stage, the other MIL loss function is imposed on the new category probabilities. In the inference phase, the threshold value is set as 0.5 to select the disease proposals from the probability map, and determine the final detection results according to the output of the contralaterally enhanced networks.

%\newpage

\section{Experiments} \label{sec:exper}

\subsection{Datasets \& Evaluation Metrics}

\noindent \textbf{Fully Supervised Dataset} We collect a private dataset to validate fully supervised disease detection methods. The dataset includes 31,000 frontal-view X-ray images which belong to 30 disease classes. 155,000 lesions indicated by bounding boxes are carefully annotated by broad-certificated radiologists. In average, there are 5 bounding boxes per image and 5,100 bounding boxes per disease category. The distributions of disease with respect to images and bounding boxes are presented in Table~\ref{tab:dataset}. The dataset is split into a training set of 27,000 images, a testing set of 2,100 images, and a validation set of 1,900 images. Three metrics based on average precision (AP) are utilized to evaluate the disease detection methods.
\begin{itemize}
\item[1.] AP-center: If the center of the predicted box is located inside certain ground-truth box having the same disease category, the box is a true positive; otherwise, it is a false positive.
\item[2.] AP50: If the IoU between the predicted box and the ground-truth box is larger than 50\%, the box is a true positive; otherwise, it is a false positive.
\item[3.] AP75: If the IoU between the predicted box and the ground-truth box is larger than 75\%, the box is a true positive; otherwise, it is a false positive.
\end{itemize}

\begin{table}[t]\setlength{\tabcolsep}{3pt}
    \centering
    %\scalebox{0.8}{%
    \begin{tabular}{l|l|ccc|cc}
        \toprule
        Framework & Version & AP-center & AP50 & AP75 & Para(MB) & FPS \\ \hline
        \multirow{2}{*}{Faster R-CNN C4} & Vanilla & 35.76 & 26.11 & 7.57 & 130 & 7.0 \\
        & Ours & \textbf{39.40} & \textbf{29.17} & \textbf{9.82} & 160 & 5.1\\ \hline
        \multirow{2}{*}{Faster R-CNN FPN} & Vanilla & 37.57 & 28.20 & 8.60 &160 &9.5 \\
        & Ours & \textbf{40.01} & \textbf{32.00} & \textbf{11.12} & 210 &7.5 \\ \hline
        \multirow{2}{*}{RetinaNet} & Vanilla & 38.02 & 28.93 & 10.01  &145 & 9.2 \\
        & Ours & \textbf{40.91} & \textbf{32.16} & \textbf{11.87} & 195& 7.1\\ \hline
        \multirow{2}{*}{CenterNet} & Vanilla & 38.19 & 29.02 & 10.11 & 800 & 3.7\\
        & Ours & \textbf{41.02} & \textbf{33.17} & \textbf{12.34} & 830 & 2.9 \\
        \bottomrule
    \end{tabular}%}
    \caption{Comparison with baseline detection models on the fully supervised dataset. Our method can significantly improve 4 existing baseline models without losing much computation efficiency.}
     \label{base}
\end{table}

\begin{table}[t]\setlength{\tabcolsep}{5pt}
    \centering
    \scalebox{1}{%
    \begin{tabular}{|c|l|l|c|c|c|}
        \hline
         Proposal & STN & Fusion & AP-center & AP50 & AP75 \\
        \hline
        \hline
        \multirow{4}{*}{C4} & w/o & w/o & 35.76 & 26.11 & 7.57 \\
        & w/o & w/ CIA & 36.01 & 27.32& 8.83  \\
        & w   & w/ CIA & 37.20 & 28.27 & 9.46  \\
        & w/o & w/ AF  & 36.12 & 26.47 & 8.69  \\
        & w/  & w/ AF  & 37.11 & 27.14 & 9.13  \\
        & w/o & w/ ASF & 36.70 & 27.92 & 9.05  \\
        & w/  & w/ ASF & \textbf{39.40} & \textbf{29.17} & \textbf{9.82}  \\
        \hline
        \hline
        \multirow{4}{*}{FPN} & w/o & w/o & 37.57 & 28.20 & 8.60  \\
        & w/o & w/ CIA & 38.14 & 30.21 & 9.54  \\
        & w/  & w/ CIA & 39.01 & 31.17 & 10.43  \\
        & w/o & w/ AF & 38.65 & 30.11 & 9.27  \\
        & w/  & w/ AF & 39.21 & 31.28 & 10.36  \\
        & w/o & w/ ASF & 38.73 & 30.45 & 9.97  \\
        & w/  & w/ ASF & \textbf{40.01} & \textbf{32.00} & \textbf{11.12}  \\
        \hline
    \end{tabular}}
    \caption{Ablation study on the fully supervised dataset. Core components in our method, including a spatial transformer network (STN) and our proposed additive and subtractive fusion (ASF), are beneficial for identification and localization of diseases. CIA indicates the features of disease proposals and their contralateral patches are fused with the contrast induced attention \cite{liu2019align}. AF means only the addition operation is used for feature fusion.}
     \label{modules}
\end{table}

\begin{figure}[t]
\begin{center}
\includegraphics[width=1\linewidth]{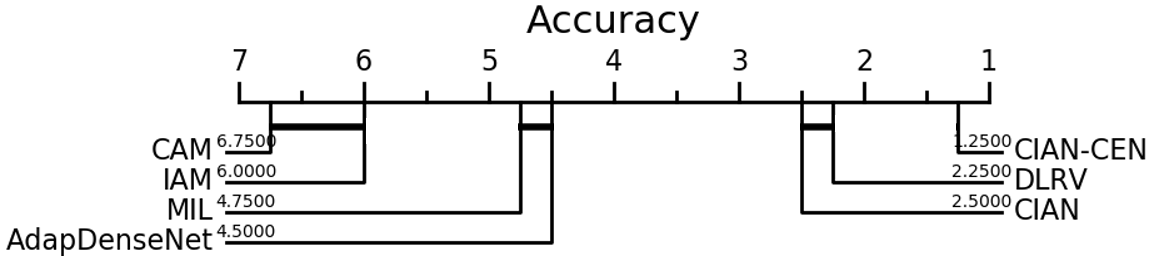}
\end{center}
\caption{We generate a critical difference diagram to make comparisons between pairs weakly supervised disease detection methods, based on the Wilcoxon-Holm method. These thick horizontal lines means the linked methods are not significantly different.}\label{fig:stat-test}
\end{figure}

\begin{figure*}[t]
\center
\includegraphics[width=0.9\linewidth]{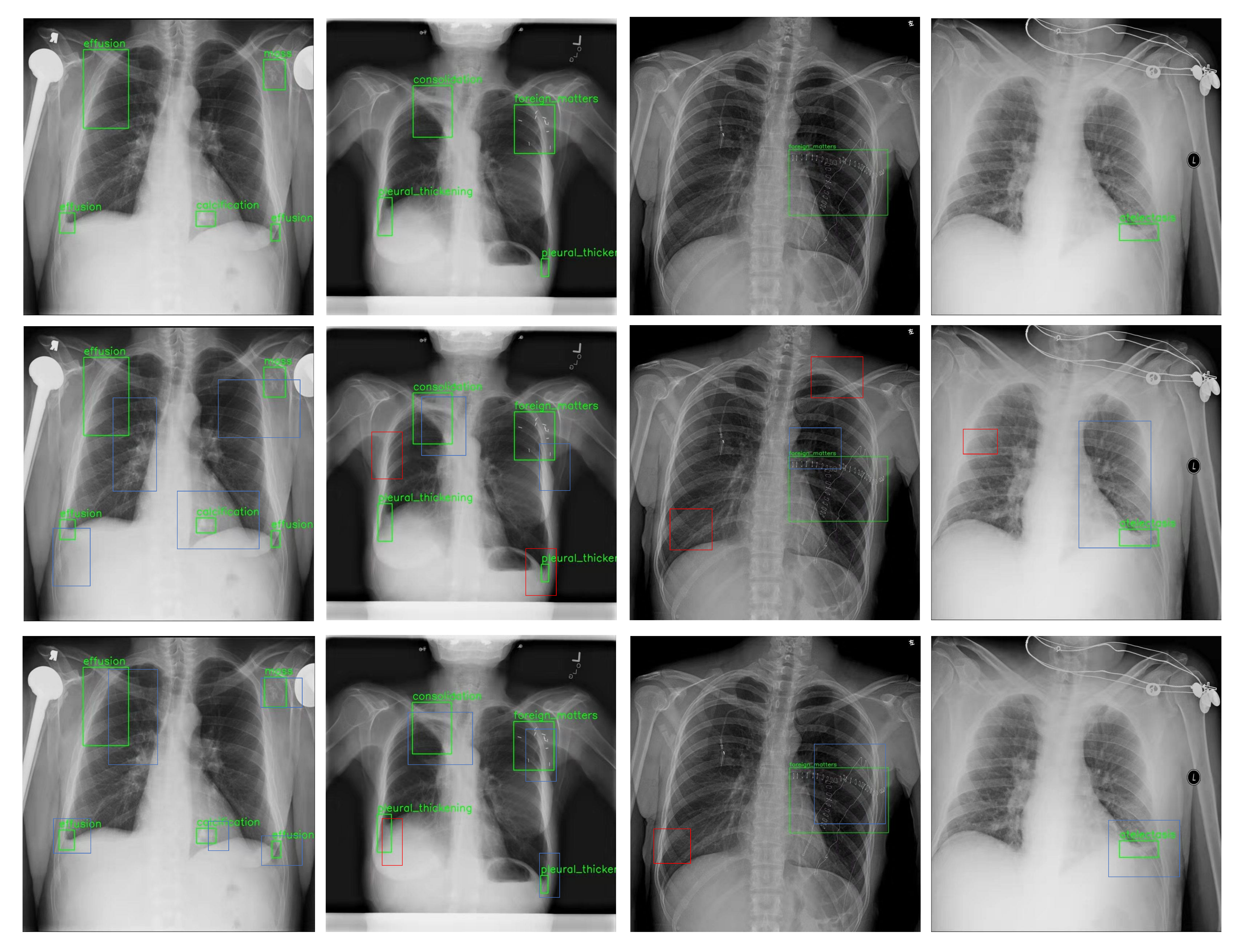}
\caption{Visualization of fully supervised disease detection. The original images, results produced by `Faster R-CNN C4' \cite{ren2015faster}, and results improved by our method are shown from top to bottom. Green, blue and red boxes represents ground-truths,  true positive predictions, and false positive predictions, respectively. We can see that our approach can produce more accurate localizations.}\label{fig:visual}
\end{figure*}

\begin{table*}[t]
    \centering
   \scalebox{1}{%
    \begin{tabular}{c|r|c|c|c|c|c|c|c|c|c}
        \toprule
        T (IoU) &  Model & Atelectasis & Cardiomegaly & Effusion & Infiltration  & Mass & Nodule & Pneumonia & Pneumothorax & Mean\\
        \midrule
        \multirow{4}{*}{0.1}
        &CAM~\cite{wang2017chestx} & 0.69 & 0.94 & 0.66 & 0.71 & 0.40 & 0.14 & 0.63 & 0.38 & 0.57 \\
        &AdapDenseNet~\cite{zhou2018weakly} & 0.43 & 0.97 & 0.69 & 0.82 & 0.55 & 0.22 & 0.78 & 0.33 & 0.60 \\
        &IAM~\cite{cai2018iterative} & 0.68 & 0.97 & 0.65 & 0.52 & 0.56 & 0.46 & 0.65 & 0.43 & 0.62 \\
        &  MIL~\cite{li2018thoracic} & 0.59 & 0.81 & 0.72 & 0.84 & 0.68 & 0.28 & 0.22 & 0.37 & 0.57 \\
        &  MIL$^\star$~\cite{li2018thoracic,liu2019align} & 0.43 & 0.82 & 0.72 & 0.72 & 0.52 & 0.42 & 0.20 & 0.70 & 0.57 \\
        &  DLRV~\cite{zhang2020thoracic} & 0.59 & 0.99 & 0.85 & 0.76 & 0.61 & 0.23 & 0.68 & 0.49 & 0.65 \\ \cline{2-11}
        &  CIAN~\cite{liu2019align}  & 0.67 & 0.86 & 0.71  & 0.83 & 0.77 & 0.45 & 0.29 & 0.40 & 0.62 \\
        &  CIAN-CEN & 0.71 & 0.91 & 0.74  & 0.85 & 0.79 & 0.47 & 0.40 & 0.50 & \textbf{0.67} \\
       \midrule
        \multirow{3}{*}{0.3}
        &CAM~\cite{wang2017chestx} & 0.24 & 0.46 & 0.30 & 0.28 & 0.15 & 0.04 & 0.17 & 0.13 & 0.22 \\
        &AdapDenseNet~\cite{zhou2018weakly} & 0.16 & 0.92 & 0.33 & 0.54 & 0.18 & 0.14 & 0.50 & 0.14 & 0.36 \\
        &IAM~\cite{cai2018iterative} & 0.33 & 0.85 & 0.34 & 0.28 & 0.33 & 0.11 & 0.39 & 0.16 & 0.35 \\
        &  MIL$^\star$~\cite{li2018thoracic,liu2019align} & 0.24 & 0.75 & 0.47 & 0.60    & 0.30  & 0.05  & 0.21 & 0.33 & 0.37  \\
        &  DLRV~\cite{zhang2020thoracic} & 0.51 & 0.96 & 0.56 & 0.67 & 0.45 & 0.16 & 0.43 & 0.21 & \textbf{0.50} \\ \cline{2-11}
        &  CIAN~\cite{liu2019align} & 0.45 & 0.73 & 0.51 & 0.74 & 0.61 & 0.19 & 0.17 & 0.26 & 0.46 \\
        &  CIAN-CEN   & 0.47 & 0.75 & 0.55  & 0.76 & 0.62 & 0.21 & 0.19 & 0.29 & 0.48 \\
       \midrule
        \multirow{3}{*}{0.5}
        &CAM~\cite{wang2017chestx} & 0.05 & 0.18 & 0.11 & 0.07 & 0.01 & 0.01 & 0.03 & 0.03 & 0.06 \\
        &AdapDenseNet~\cite{zhou2018weakly} & 0.06 & 0.78 & 0.23 & 0.28 & 0.08 & 0.14 & 0.21 & 0.05 & 0.23  \\
        &IAM~\cite{cai2018iterative} & 0.11 & 0.60 & 0.10 & 0.12 & 0.07 & 0.03 & 0.17 & 0.08 & 0.16 \\
        & MIL$^\star$~\cite{li2018thoracic,liu2019align} & 0.15 & 0.67 & 0.30  & 0.41 & 0.22 & 0.02 & 0.14 & 0.12 & 0.25 \\
        &  DLRV~\cite{zhang2020thoracic} & 0.20 & 0.92 & 0.19 & 0.39 & 0.20 & 0.06 & 0.18 & 0.04 & 0.27 \\ \cline{2-11}
        &  CIAN~\cite{liu2019align} & 0.31 & 0.65 & 0.37 & 0.59 & 0.48 & 0.07 & 0.09 & 0.20 & 0.35\\
        &  CIAN-CEN  & 0.32 & 0.68 & 0.39  & 0.61 & 0.49 & 0.07 & 0.15 & 0.21 & \textbf{0.36} \\
        \midrule
        \multirow{3}{*}{0.7}
        &CAM~\cite{wang2017chestx} & 0.01 & 0.03 & 0.02 & 0.00 & 0.00 & 0.00 & 0.01 & 0.02 & 0.01 \\
        &AdapDenseNet~\cite{zhou2018weakly} & 0.02 & 0.23 & 0.22 & 0.11 & 0.04 & 0.14 & 0.08 & 0.01 & 0.11  \\
        &IAM~\cite{cai2018iterative} & 0.01 & 0.17 & 0.01 & 0.02 & 0.01 & 0.00 & 0.02 & 0.02 & 0.03 \\
        & MIL$^\star$~\cite{li2018thoracic,liu2019align} & 0.03 & 0.57 & 0.08  & 0.04 & 0.1 & 0.02 & 0.06 & 0.10 & 0.10   \\
        &  DLRV~\cite{zhang2020thoracic} & 0.04 & 0.72 & 0.03 & 0.15 & 0.02 & 0.00 & 0.03 & 0.01 & 0.13 \\ \cline{2-11}
        &  CIAN~\cite{liu2019align}  & 0.11 & 0.53 &  0.18 & 0.27 & 0.26 & 0.03 & 0.04 & 0.16 & 0.19 \\
        &  CIAN-CEN  & 0.13 & 0.54 & 0.19  & 0.27 & 0.27 & 0.04 & 0.05 & 0.17 & \textbf{0.21} \\
        \bottomrule
    \end{tabular}}
    \caption{Comparison with other weakly supervised disease detection methods on NIH chest X-ray dataset. T(IoU) means the threshold value of IoU used to match predicted results and ground-truths.
     Disease localization accuracy are evaluated at T(IoU)-s in \{0.1, 0.3, 0.5, 0.7\}.
     `MIL$^\star$' indicates the results of~\cite{li2018thoracic} which are re-implemented by~\cite{liu2019align}.}
     \label{tab:weak}
\end{table*}

\noindent \textbf{Weakly Supervised Dataset} The NIH chest X-ray dataset \cite{wang2017chestx} is used for weakly supervised disease detection. It contains 112,120 frontal-view X-ray images with 14 disease classes. Bounding box annotations are provided for 880 images. In this paper, we use images with class annotations during the training stage, while the 880 images with bounding box annotations are used for testing. We evaluate the performance of disease detection, following the metrics in~\cite{wang2017chestx,li2018thoracic,liu2019align}. The threshold of IoU for identifying true positive detections varies from 0.1 to 0.7, in step of 0.2.

\subsection{Fully Supervised Disease Detection} \label{sec:exper-full}

\noindent \textbf{Spine Segmentation} For training the spine segmentation network, the spine regions of 8000 images are carefully annotated by radiologists. These images are randomly split into 10 folds. Cross validation is conducted to validate the segmentation performance. The Dice similarity coefficient (DSC) and three other metrics including pixel accuray (P.ACC), mean accuracy (M.ACC), mean IU (M.IU) and frequency weighted IU (F.W.IU) proposed in~\cite{long2015fully}  are used for evaluation. The experimental results are presented in Table~\ref{tab:spineseg}

%the results are presented in Table~\ref{tab:spineseg}.
%We firstly trained a spine segmentation network to extract the spine mask of chest X-ray images. Specifically, about 8000 chest X-ray images are carefully annotated by senior medical experts. In addition, we adopted PSPNet [] as our segmentation methods. We used the same setting as illustrated in [] and only modified its class num to suitable our task. In detail, the experimental results of our spine segmentation network are shown in Table~\ref{tab:spineseg}

\begin{table}[t]
\centering
\setlength\tabcolsep{12pt}
\begin{tabular}{ccccc}
    \toprule
        DICE   & P.ACC  & M.ACC  & M.IU   & F.W.IU \\ \hline
        96.4\% & 97.1\% & 80.4\% & 64.2\% & 89.2\% \\
    \bottomrule
    \end{tabular}
\caption{The performance of PSPNet in the spine segmentation task. Dice similarity coefficient (DSC), pixel accuracy (P.ACC), mean accuracy (M.ACC), mean IU (M.IU), frequency weighted IU (F.W.IU) are used for evaluation.}
\label{tab:spineseg}
\end{table}

\color{black}{\noindent \textbf{Disease Detection} The four baseline models, including `Faster R-CNN C4', `Faster R-CNN FPN', RetinaNet and CenterNet, are re-implemented using our fully supervised dataset. We attempt to incorporate our proposed module into all of them as introduced in Section~\ref{sec:fsd}. The quantitative comparisons are presented in Table~\ref{base}.
The proposed module can significantly improve the four baseline models. The improvement of AP-center brought by our method is 3.64\% (from 35.76\% to 39.40\%) or 2.44\% (from 37.57\% to 40.01\%), when employing `Faster RCNN C4' or `Faster RCNN FPN' to produce disease proposals.
Under metrics AP50 and AP75, our method can also outperform all compared methods. For example, the variant of our method equipped with `Faster R-CNN C4' achieves 29.17\% AP50 and 9.82\% AP75 which is 3.06 and 2.25 higher than the results of original `Faster R-CNN C4' respectively. The improvements brought by our method are 3.80 (AP50) and 2.52 (AP75) when using the baseline of `Faster R-CNN FPN'.}
The confusion matrices \footnote{\url{https://github.com/kaanakan/object_detection_confusion_matrix}} of `Faster R-CNN FPN'and its variant improved by our method are presented in Fig. \ref{fig:confusion}.

Results of disease detection in 4 CXRs are visualized in Fig.~\ref{fig:visual}.
From top to down are original images,  results of the original `Faster R-CNN C4', and results of the variant of `Faster R-CNN C4' which is improved by our module. The green, blue and red boxes stand for ground-truths, true positives and false positives, respectively.
The experimental results demonstrate that the contralateral context information extracted by our proposed module is beneficial to the detection of diseases in CXRs.

%The baseline `Faster RCNN C4' and `Faster RCNN FPN' process 7.0 and 9.5 images per second, respectively, while the counterparts of these models which are modified with our proposed module process 5.1 and 7.5 images per second, respectively. Our method is capable of improving the baseline detection models without losing much efficiency.

\begin{figure*}[t]
\begin{center}
\includegraphics[width=1\linewidth]{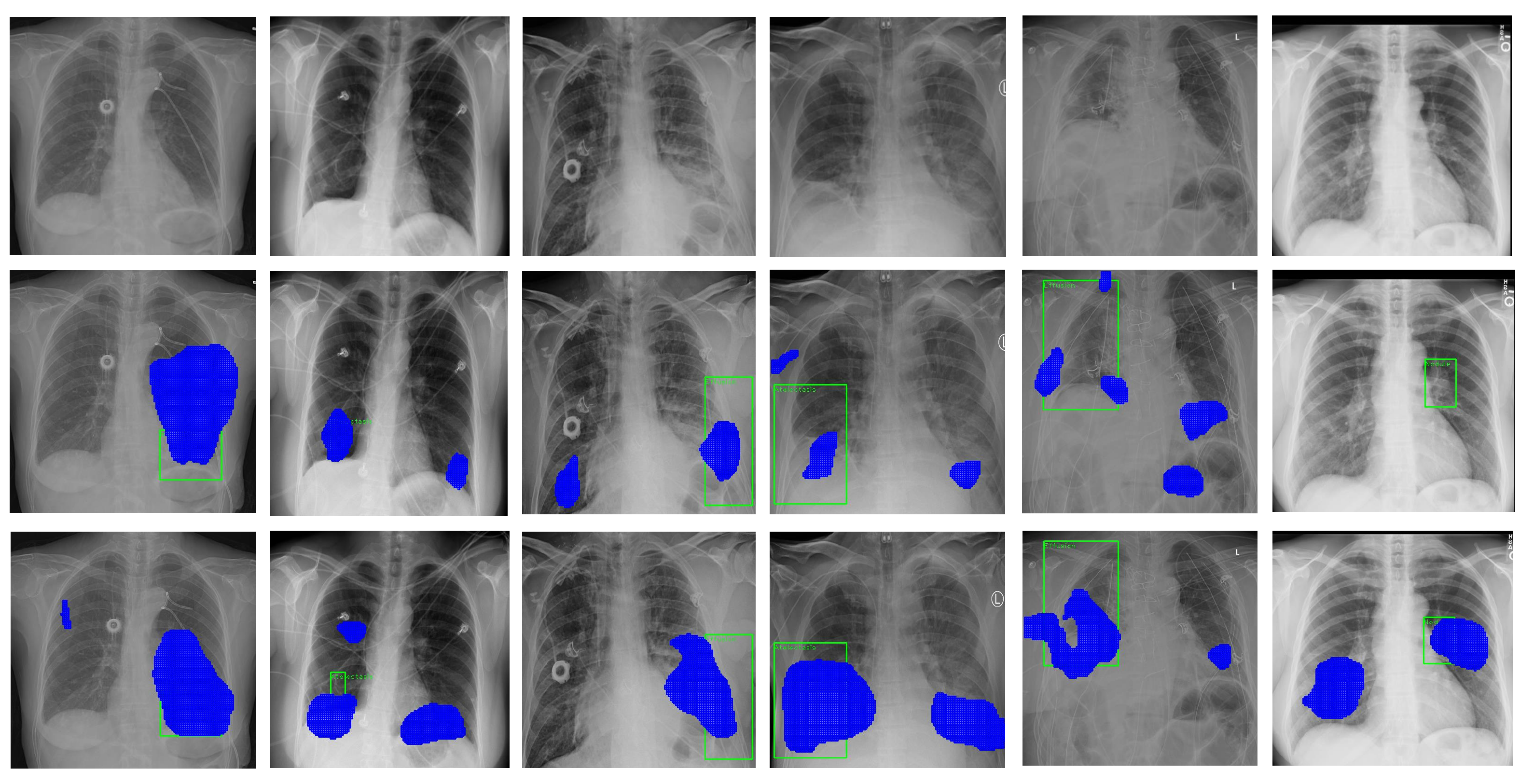}
\end{center}
\caption{Visualization of weakly supervised disease detection. The original images are presented in the first row. Results produced by \cite{liu2019align} and our method are visualized in the second and third row respectively. Green boxes stand for ground-truths and blue regions indicate predicted results. We can see that our approach can output more accurate localization results.}\label{resultsvis3}
\end{figure*}

\subsection{Weakly Supervised Disease Detection}
In this task, we adopt images with only image-level labels for training, and images with both bounding box and class annotations for testing. We compare our method against various existing methods proposed in~\cite{wang2017chestx,li2018thoracic,zhou2018weakly,cai2018iterative,liu2019align,zhang2020thoracic}.
As shown in Table~\ref{tab:weak}, our method achieves the best performance in overall. Compared to the baseline model~\cite{liu2019align}, our method achieves consistently higher accuracy under all IoU thresholds. For example, in case of using 0.3 as IoU threshold, it produces results with accuracy of 0.48, surpassing~\cite{liu2019align} by 0.02. When threshold of IoU is set as 0.5 and 0.7, our approach achieves accuracy of 0.36 and 0.21, with a lead of 0.01 and 0.02 over~\cite{liu2019align}, respectively. In Fig.~\ref{resultsvis3}, we illustrate disease detection results of several cases under the condition of the weak supervision. From top to down are original images, results of~\cite{liu2019align}, and results produced by our method. The green boxes stand for ground-truths and blue regions indicate predictions inferred by disease localization models. The results verify the effectiveness of our method in weakly supervised disease detection. Following \cite{fawaz2019deep}, we generate a critical difference diagram for comparing our method against other weakly supervised methods, based on the Wilcoxon-Holm method \footnote{\url{https://github.com/hfawaz/cd-diagram/}}. The mean accuracies under different IoUs are used as the input. As shown in Fig.~\ref{fig:stat-test}, our proposed method achieves the best rank.

\begin{figure*}[t]
\begin{center}
\includegraphics[width=1\linewidth]{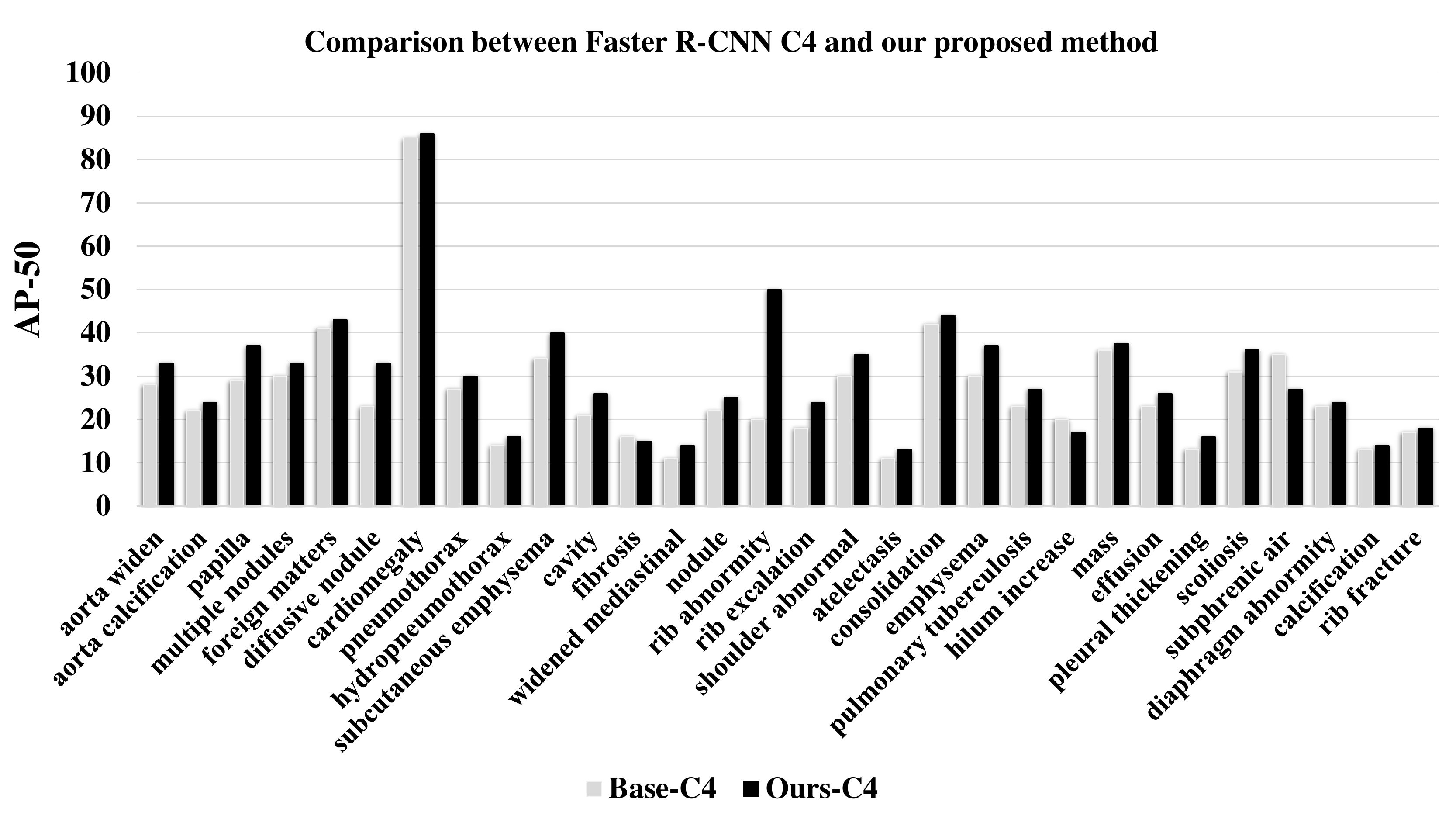}
\end{center}
\caption{The AP50 metric values of 30 disease categories. `Faster R-CNN C4' is used to extract disease proposals in our method. The incorporation of contralateral patches is beneficial for improving the detection performance of most diseases.}\label{fig:classes}
\end{figure*}

\iffalse
\begin{table}[t]
    \centering
    \scalebox{1.5}{%
    \begin{tabular}{cc}
        \toprule
         statistic & p-value \\
        \midrule
        41.5 & 0.00039  \\
        \bottomrule
    \end{tabular}}
    \caption{The Wilcoxon signed-rank test tests the null hypothesis that two results (Our Faster-RCNN C4 vs. Base Faster-RCNN C4). In particular, it tests whether the distribution of the differences between (Our Faster-RCNN C4 vs. Base Faster-RCNN C4) is symmetric about zero. It is a non-parametric version of the paired T-test.}
     \label{stat-class-disease}
\end{table}
\fi
\subsection{Ablation Study}
Ablation studies are conducted to discuss the effectiveness of core modules in our method. Both `Faster R-CNN C4' and `Faster R-CNN FPN' are used as baseline models. To validate the effectiveness of the STN, a variant of our method without STN, which uses the preliminary contralateral patches to enhance the features of disease proposals, is implemented. In contrast to our proposed additive and subtractive fusion (ASF) strategy, we also apply a counterpart of our feature module with only additive operation (AF) or the contrast induced attention (CIA)~\cite{liu2019align} which is based on the subtractive operation to fuse features of each disease proposal and its contralateral patch. As we can see in~\ref{modules}, compared to the baseline models, the usage of preliminary contralateral patches can give rise to marginally better performance. The adoption of STN is capable of improving all detection evaluation metrics in all settings. For example, the value of AP-center is promoted from 36.70 to 39.40, once STN is adopted in the variant of `Faster R-CNN C4' with ASF. For feature fusion strategies, our proposed ASF outperforms CIA under all conditions. For example, ASF induces detection results with 2.20 (39.40 vs 37.20) or 1.00 (40.01 vs 39.01) higher AP-center than the results of CIA when using `Faster R-CNN C4' or `Faster R-CNN FPN' as the baseline model, respectively. Besides, the detection performance is degraded without using any of the additive and subtractive operations. It indicates that the two operations can complement each other.
In summary, exhaustive ablation studies demonstrate the superiorities of core modules devised in this paper.

\subsection{Performance in Individual Disease Categories}
\color{black}{The AP50 metric values of 30 disease categories produced by `Faster R-CNN C4' and our proposed method are presented in Fig.~\ref{fig:classes}. Compared with the baseline `Faster R-CNN C4', our proposed method achieves better results in  the detection of most disease categories. Our method is particularly advantageous at detecting abnormalities residing in structures having similar counterparts in the contralateral side of the chest,  such as lungs and ribs. Especially, the AP-50 of rib abnormality detection is improved from around 20\% to 50\% after incorporating our method into `Faster R-CNN C4'.
For diseases of organs which only exist in single half of the chest, such as cardiomegaly, our method can still improve the identification performance, since the contralateral context information can also be used to ignore overlapping or surrounding distortion signals.}

For statistical analysis, the Wilcoxon signed-rank test \footnote{\url{https://docs.scipy.org/doc/scipy/reference/generated/scipy.stats.wilcoxon.html}} is conducted to compare the distributions of AP50 across diseases, produced by the original `Faster R-CNN C4' and its improved variant. The sum of the ranks of the differences is 41.5, and the p-value for the test is 0.00039.  Hence, we would reject the null hypothesis that the two groups of AP50 are from the same distribution under a confidence level of 5\%.

The above experiments demonstrate that, the retrieved contralateral patches are able to strengthen the feature representations of proposals when identifying and locating most disease lesions.
%\new{Tabel~\ref{stat-class-disease} shows the difference between two methods statistically significant in terms of disease categories. Hence, we would reject the null hypothesis at a confidence level of 5\%, concluding that there is a difference in category accuracy between the groups. To confirm that the median of the differences can be assumed to be positive}

\section{Conclusion}
We propose a novel module, \textit{Contralaterally Enhanced Networks}, for disease localization in chest X-ray images.
Our method aims at taking advantage of the thoracic contralateral context information to enhance the feature representations of disease proposals. The spine line is regarded as the symmetry axis to obtain an initial contralateral patch for each disease proposal. Then a spatial transformer network is devised to refine the pose of the initial contralateral patch. The disease proposal and its retrieved contralateral patch are  fused to predict final disease classification and localization. Experiments on a carefully annotated dataset demonstrate our proposed module improves existing two-stage and one-stage detection methods, Faster R-CNN C4, Faster R-CNN RPN, RetinaNet and CenterNet, by 3.06, 3,80, 3.23 and 4.15 on the AP50 metric, respectively.
Our method can also be applied in weakly supervised disease localization and achieves state-of-the-art performance on the NIH chest X-ray dataset.

\section{Future Work}
As discussed above, the contralaterally enhanced networks can effectively improve the disease performance in both fully and weakly supervised disease localization in chest X-ray images. The limitations of our method are as follows: 1) The retrieval of the contralateral patches is dependent to the localization of the spine line; 2) The disease regions neglected by the proposal extraction model can not be recovered. To settle these issues, it deserves further study to design more efficient module to explore the contralateral contextual information. Furthermore, the replication of small thoracic structures exists in the same side of the chest. Thus, how to exploit such relationships for disease detection is also an interesting research topic.
\bibliographystyle{IEEEtran}
\bibliography{journalbib}

\end{document}